\documentclass[journal]{IEEEtran}

\usepackage{graphicx}
\usepackage{times}   % assumes new font selection scheme installed
\usepackage{amsmath} % assumes amsmath package installed
\usepackage{amssymb} % assumes amsmath package installed
\usepackage[utf8]{inputenc}
\usepackage[english]{babel}
\usepackage[linesnumbered,ruled,vlined]{algorithm2e}
\usepackage{svg}
\usepackage{pdftexcmds}

\setsvg{
	inkscape = inkscape -z -D % conversion options for svg package, export drawing instead of page
}

\newtheorem{theorem}{Theorem}

\newtheorem{lemma}[theorem]{Lemma}

\title{On the Design of an Insurance Mechanism for \\ Reliability Differentiation in Electricity Markets}
\author{Farhad~Billimoria,~\IEEEmembership{Member,~IEEE,}
	Filiberto~Fele,
	Iacopo~Savelli,
	Thomas~Morstyn,~\IEEEmembership{Member,~IEEE,}
	and~Malcolm~McCulloch, ~\IEEEmembership{Senior Member,~IEEE}% <-this % stops a space
	\thanks{Corresponding author: F. Billimoria is with the Energy \& Power Group, Department
		of Engineering Science, University of Oxford, Oxford OX1 2JD,
		United Kingdom e-mail: farhad.billimoria@eng.ox.ac.uk.}% <-this % stops a space
	\thanks{F. Fele (filiberto.fele@eng.ox.ac.uk) and M. McCulloch (malcolm.mcculloch@eng.ox.ac.uk) are with the Energy \& Power Group, Department of Engineering Science, University of Oxford, UK}
	\thanks{I. Savelli (iacopo.savelli@smithschool.ox.ac.uk) is with the Smith School of Enterprise and the Environment, University of Oxford, UK}
	\thanks{T. Morstyn (thomas.morstyn@ed.ac.uk) is with the School of Engineering, University of Edinburgh, UK}% <-this % stops a space
	\thanks{\textbf{This work has been submitted to the IEEE for possible publication. Copyright may be transferred without notice, after which this version may no longer be accessible.}}% <-this % stops a space

	}
\begin{document}
\maketitle
\thispagestyle{empty}
\pagestyle{empty}
%%%%%%%%%%%%%%%%%%%%%%%%%%%%%%%%%%%%%%%%%%%%%%%%%%%%%%%%%%%%%%%%%%%%%%%%%%%%%%%%

\begin{abstract}
Securing an adequate supply of dispatchable resources is critical for keeping a power system reliable under high penetrations of variable generation. Traditional resource adequacy mechanisms are poorly suited to exploiting the growing flexibility and heterogeneity of load enabled by advancements in distributed resource and control technology. To address these challenges this paper develops a resource adequacy mechanism for the electricity sector utilising insurance risk management frameworks that is adapted to a future with variable generation and flexible demand. The proposed design introduces a central insurance scheme with prudential requirements that align diverse consumer reliability preferences with the financial objectives of an insurer-of-last-resort. We illustrate the benefits of the scheme in (i) differentiating load by usage to enable better management of the system during times of extreme scarcity, (ii) incentivising incremental investment in generation infrastructure that is aligned with consumer reliability preferences and (iii) improving overall reliability outcomes for consumers.
\end{abstract}
%%%%%%%%%%%%%%%%%%%%%%%%%%%%%%%%%%%%%%%%%%%%%%%%%%%%%%%%%%%%%%%%%%%%%%%%%%%%%%%%
\section{INTRODUCTION}
In this paper we posit that  the \textit{missing money problem} in electricity markets can be addressed by a reliability insurance scheme that enables differentiation and prioritisation of demand. Resource adequacy is particularly relevant today as the de-carbonisation of the electricity sector requires the deployment of large amounts of variable renewable energy (VRE), which is expected to supply 70-90\% of global electricity demand by 2050 \cite{IPCC2018}. An electricity system with large penetrations of VRE will require an adequate capacity of dispatchable resources to balance periods of intermittent or low renewable resource availability \cite{Simshauser2018}.

In most liberalised markets, electricity is typically dispatched in economic merit-order and cleared on the basis of a marginal price \cite{Simshauser2018}. In theory, the marginal price is capable of stimulating generation investment to ensure long-term generation capacity adequacy \cite{Schweppe1988a}. However, in practice a range of factors, including system operator interventions and administrative caps on market prices, restrict power prices from reaching the theoretical value of lost load (VOLL) \cite{Hogan2013}. This leads to the well-studied \textit{missing money problem} where generators face a chronic shortage of revenue, while retailers avoid the full financial impact of lost load, resulting in under-hedging of load and underinvestment in resource capacity \cite{Newbery2018}. This becomes increasingly challenging as markets become dominated by variable resources with zero or negligible short-run marginal costs \cite{Newbery2018}. While some regions have persisted with an \textit{energy-only market} design\footnote{An energy-only market design, is one where the centrally cleared marginal price is the sole source of market revenue available to generators. No other remuneration mechanisms (e.g. capacity markets) are present though participants are free to contract outside of the central spot market. Regions with this design include New Zealand, the Electricity Reliability Council of Texas and the National Electricity Market of Australia},  many (including the UK, Germany and Belgium) have recently introduced centralised resource adequacy mechanisms to support dispatchable resource investment \cite{Billimoria2019}.

However, traditional resource adequacy mechanisms, such as capacity markets, are poorly suited to the changing role of load in balancing the system \cite{Bushnell2017}. The advent of distributed energy and storage technology, and enhanced load controllability has enabled greater differentiation in consumer preferences for electricity service \cite{Bose2019}. For example, consider the benefits that could be obtained, in scarcity situations, from prioritising essential uses of electricity (including where electricity is a gate to other essential services such as heat and water) over other uses which have little impacts on overall quality of service \cite{Gils2014,Busby2021}. As capacity markets tend to rely on standardised metrics (such as reserve margins or reliability demand curves) they are less capable of reflecting the increasingly heterogeneous and flexible nature of load in the power system \cite{Lambin2020}. Gottstein and Skillings \cite{Gottstein2012} encourage our \textit{"collective thinking... to evolve beyond capacity markets"} to address the reliability challenges of a decarbonized grid. 

We take a fresh look at the design of resource adequacy mechanisms by introducing insurance risk management and loss provisioning techniques to the assessment of electricity reliability. The microeconomic model of an insurer is as a manager of tail risk \cite{Kaas2008}. Tail risk relates to financial loss exposures from extreme or low-probability outcomes (i.e. the so-called tail of a probability distribution). This suggests a natural applicability  to assessment of resource adequacy in power systems. The scope of the paper is as follows (i) the design of a centralised insurance mechanism and the interaction with an operational scheme for priority curtailment of load; (ii) the design of a decision making framework for the central insurer; and (iii) a comparison of lost load and cost outcomes of the insurance-based design against an energy-only market design. We are focused in this paper on generation capacity expansion only and do not consider network investment at this stage.

The theory of reliability insurance was originally proposed in \cite{Hung-PoChaoandRobertWilsonSource2016} as a contractual mechanism to price the risk of differentiated reliability and to remediate the reliability externality effects associated with energy-only market designs \cite{Kiesling2009}. Reliability insurance offers consumers compensation for electricity interruptions in return for an upfront premium \cite{Oren1990}. This in turn creates an incentive for the insurance counterparty to mitigate interruption risk through portfolio diversification and investment in or contracting with generators \cite{Chao1988}. Chao in \cite{Chao2012} extends the original model to a two-stage model comprising a forward subscription and real-time consumption stage. In \cite{Fuentes2020} an agent-based model is used to confirm the viability of an insurance market that allows consumers to efficiently transfer risk to the utility, based on willingness to pay and risk profiles. In both centralised and decentralised designs for reliability insurance, an operational priority curtailment scheme is required to mitigate free-rider effects, which occur when a consumer chooses not to insure itself but nevertheless obtains reliability benefits from additional investment in the system \cite{Billimoria2019}. 

A related stream of work is the design of priority service schemes. While these works eschew consideration of an insurance pricing framework for risk assessment, they remain relevant in informing the design of priority curtailment  schemes which could be linked to an insurance mechanism. Under a priority service scheme, electricity is considered a service that can be offered with different levels of reliability \cite{Mou2020}. As described in \cite{Gerard2019} residential consumers are offered a menu of price–reliability pairs with contracts being created for \textit{strips of power} with a particular reliability. A simple colour-tagging scheme (e.g. a 'red','yellow','green' traffic light system) allows consumers to designate devices/appliances to strips with an energy router communicating those preferences to a system operator \cite{Papalexopoulos2013}. The utility can then implement reliability differentiation on the supply side via tolling and availability contracts with generators \cite{Woo2019,Woo2019a}. 

In this work we seek to address two gaps in the research. First, despite recent interest in the concept of reliability insurance there has been relatively little focus on the decision-making model of the reliability insurer itself. In other words, how does the insurer make contracting and investment decisions in a real-world setting? The literature on reliability insurance has to date relied upon a simplified model of a risk-averse consumer and a risk-neutral utility to justify the case for efficient risk-transfer via insurance \cite{Hung-PoChaoandRobertWilsonSource2016,Oren1990,Fuentes2020}. While this can usefully illustrate the theoretical rationale for reliability insurance, it does not allow a practical appreciation of how an insurer would make decisions given constraints on funding, capital availability and risk. The second gap relates to the integration of a reliability insurer within an electricity market. Existing work on reliability insurance and priority service assumes that the provider of insurance is a vertically integrated utility that owns all generation and supplies all consumers \cite{Fuentes2020,Mou2020}. However, many electricity markets are now liberalised and generation and supply is often competitive and disaggregated \cite{Simshauser2018}. 

To address these two gaps we propose a new \textit{energy plus insurance} market design that incorporates a centralised reliability insurance scheme as an overlay on the electricity spot market. Furthermore we develop a reliability risk metric that draws upon the body of literature in insurance theory and tail risk management. This allows for the formulation of a decision making model of the insurer that is integrated with a broader market. In doing so the paper seeks to advance from a theoretical notion of reliability insurance towards a practical construct capable of implementation. The key features of this paper are as follows:

\begin{enumerate}
	\item a novel decision making model for the reliability insurer framed as an optimization problem.
	\item a new insurance loss provisioning framework that requires the reliability insurer to assess, provision and protect against tail risks in the electricity system.
	\item the formulation of an algorithm modelling the interactions between the decisions of a centralised reliability insurer and that of independent generators in the market.
\end{enumerate}

The rest of this paper is organized as follows.  In Section II we begin with an architecture of our proposed \textit{energy plus insurance} market design, and then in Section III formulate an insurance provisioning metric that relates reliability outcomes to insurer financial risk. Using the metric, in Section IV we formalise in mathematical terms the decision making problem of the centralised insurer and outline an algorithm to compute an equilibrium (if it exists) between participants in the electricity spot market and the centralised insurer. In Section V we apply the design to a case study and present results. Section VI concludes with policy implications and extensions.

\begin{figure*}[thpb]
	\centering
		\includegraphics[height=0.80\columnwidth]{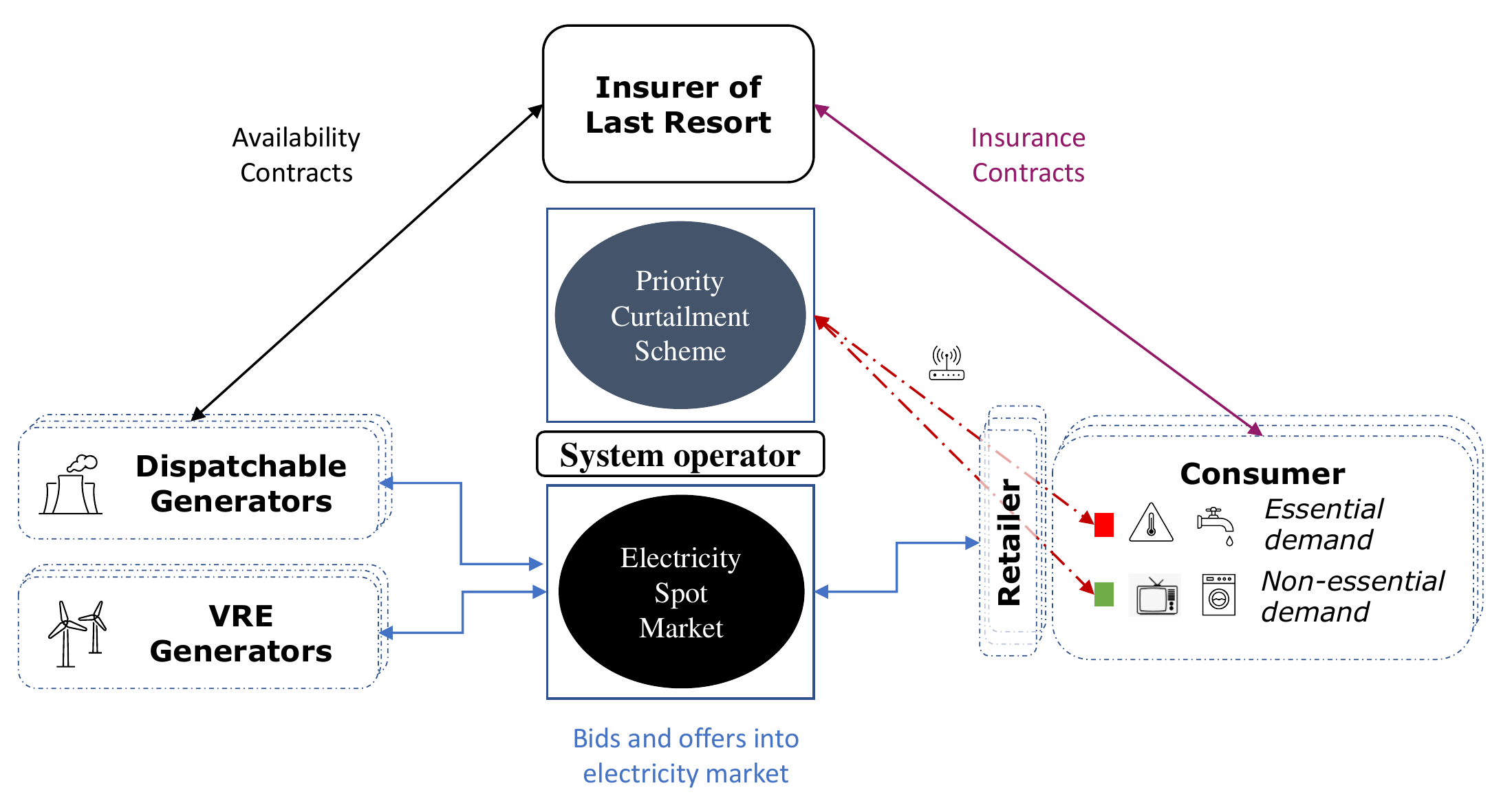}
%	\begin{tabular}{cc}     
%	\end{tabular}
	\caption{Schematic of the market architecture incorporating a centrally dispatched spot market settled on marginal prices (managed by the system operator), a priority curtailment scheme managed by the system operator, and a centralised reliability insurance scheme managed by the insurer of last resort (IOLR).  Blue arrows represent offers and bids by generation and demand respectively into the spot market.  Purple arrows represent reliability insurance contracts executed between consumers and the IOLR. Black arrows represent availability contracts executed between dispatchable generators and the IOLR. Dashed red arrows represent communication links between the system operator and load enabling priority curtailment of load in order of compensation value.}
	\label{fig:market_design}
\end{figure*}

\section{An "Energy plus Insurance" Market Design}

In this section we describe the architecture of the \textit{energy plus insurance} market design. There are three components to the proposed design (i) a centrally cleared electricity spot market managed by a system operator; (ii) a centralised reliability insurance scheme; and (iii) an operational priority curtailment scheme managed by the system operator. 

We begin with the electricity spot market. Generators offer into a gross pool, are dispatched in economic merit order, and are settled at marginal prices with an administrative market price cap limiting the price \cite{Hogan2013}. Demand is contracted to electricity retailers, but can also participate and bid directly into the spot market (though this is typically limited to large loads). Complementary administrative mechanisms include generator offer caps and market power mitigation processes \cite{Hogan2013}. In the absence of any other sources of revenue for generators, this market design is the \textit{energy only market} referred to above.

Two additional components are added to the market design.  The first is a centralised insurance scheme managed by a new entity termed the insurer-of-last-resort  (IOLR) \cite{Billimoria2019}. The IOLR is a central authority that offers electricity interruption insurance to electricity consumers and is responsible for managing the insurance scheme. The role of the IOLR can be housed in a new agency, or delegated to existing parties in the system (such as the transmission system operator). Energy consumers express their reliability preferences by voluntarily entering into reliability insurance contracts with the IOLR. In this paper we use a relatively simple formulation of an insurance contract which specifies an upfront premium charged on peak demand (in \$ per MW-peak), and a compensation value (in \$ per MWh) that is payable if the load is interrupted. The compensation value should be representative of the VOLL of the particular use of electricity. Customers can specify different compensation values that attach to different \textit{strips of power}.  Differential reliability is enabled by the compensation values that customers ascribe to different devices or uses of electricity. More complex price-menu designs that specify reliability levels and maximum coverages are possible, though this is outside of the scope of this paper.  

The third component of the design is an operational demand priority curtailment  scheme administered by the system operator \cite{Chao2012}. The compensation value specified in reliability insurance contracts would be provided to the system operator. Load across the system would be ranked in order of compensation value. In operational situations where available generation is insufficient to balance the system, load would be curtailed in order of lowest compensation value to highest compensation value. Thus under scarcity, load would be \textit{triaged} and low value or non-essential load would be curtailed first. This would be actuated through real-time communications infrastructure such as an energy router connected to the home \cite{Papalexopoulos2013}.

The IOLR is incentivised to take action to reduce its risk of paying out insurance compensation. It can do so by executing availability contracts with dispatchable generation resources, if the improvement in reliability from having those resources available offsets the cost of the availability contract. Generators may continue to invest on the basis of spot market prices alone (we call these independent generators) or may seek availability contracts with the IOLR (we call these IOLR-supported generators). A high level block diagram of the proposed market architecture is provided in Fig. \ref{fig:market_design}.  

The IOLR is funded by debt and shareholder equity, with assets invested in an investment pool. The IOLR holds assets sufficient to pay claims, and seeks to generate profits, which provides a return, compensating investors for the risk borne \cite{Rees2008}. The IOLR's profit comprises revenues from its investment pool and from insurance premiums minus its expenses.  An IOLR's expenses comprise compensation payouts from insurance portfolio, interest and principal payments on loans or other debt financing (collectively debt service payments) and availability payments to generators. International estimates of equity returns for general insurance companies range between 5-8\% depending upon risk exposures \cite{Barinov2020}. 

\section{Insurance Loss Reserving}

In this section we describe the principles governing the viability and solvency of an insurer, and use these principles to formalise prudential metrics that guide an IOLR's decision-making framework. 

A key principle of the insurance business model relates to the reserving of capital. An insurer will seek to set premiums for insuring a particular risk at a value that is in excess of the expected value of the loss claims payable from that risk. This is known as the \textit{expected value premium principle} \cite{Kaas2008}. In order to maintain solvency the insurer must also provision for potential financial losses from tail risk outcomes. This means the insurer must carry reserves against the possibility that the aggregate value of loss claims from that risk will exceed its premium income \cite{Rees2008}. These are often termed technical or insurance reserves and held in cash or equivalently secure and liquid investments. The quantity of reserves held by the insurer are sized based on applying a risk measure to profits for a given confidence level, and are typically guided by best-practice prudential risk standards and industry regulation \cite{Rees1999}. The IOLR is similarly required to maintain reserves that are sufficient to remain solvent given a portfolio of reliability insurance contracts with electricity consumers. In this paper the conditional value-at-risk (CVAR) risk measure is used given its coherency properties \cite{Rockafellar2002} and that it accounts for the shape of the tail of the probability distribution \cite{Pineda2010}. 
\begin{align}
	\tilde{\mathit{c}}^i = z^i - \frac{1}{1-\alpha^i}\sum_{\omega\in\Omega} \pi_\omega \varrho^i_{\omega} \label{cvarcon} \\
	z^i - \Psi^i_{\omega} \leq \varrho^i_{\omega},	\ \forall \omega \in \Omega \label{cvarcon2} \\
	\varrho^i_{\omega} \geq 0,	\ \forall \omega \in \Omega \label{cvarcon3}
\end{align}
The scenario-based formulation of the IOLR's CVAR ($\tilde{\mathit{c}}^i$) is defined in (\ref{cvarcon})--(\ref{cvarcon3}) where $\Psi^i_{\omega}$ represents the IOLR's net profits for the set of scenarios $\omega \in \Omega$, $z^i$ represents the value-at-risk (VAR), $\alpha^i$ represents the confidence level, $\pi_\omega$ represents the scenario probability, $\varrho^i_{\omega}$ represents the positive deviation between scenario profits and the VAR \cite{Pineda2010}. We superscript relevant variables and parameters by $i$ to distinguish between the CVAR of the IOLR and that of independent generators. 
\begin{align}
 	\tilde{\mathit{c}}^i \geq -\phi^i \label{rm}
\end{align}
We propose a prudential risk constraint (\ref{rm}) for the IOLR which requires technical reserves $\phi^i$ exceed the negative CVAR. This can be interpreted as requiring the IOLR to have reserves that cover average worst-case outcomes beyond the confidence level $\alpha^i$. Confidence levels for insurers are generally set very high to account for tail risk outcomes.  For example, the European Solvency II insurer financial risk framework requires insurers to assess risks at 99.5\% confidence level \cite{Eling2007}.  Prudent insurance risk management requires this metric must be met by IOLR.

\section{Problem formulation}

This section proposes mathematical formulations for the decision-making model of the IOLR, incorporating the prudential risk constraint above, and the decision-making model of independent generators in the market. Consequently it proposes an algorithm to model the interactions between the IOLR and independent generators, and to find an equilibrium.  

\subsection{Decision-making framework for the Insurer of Last Resort }
First, the formulation of the decision-making framework for an IOLR is set out. At a high-level, the framework takes as inputs parameters relating to electricity dispatch outcomes and consumer reliability and premium preferences, and makes decisions regarding the execution of reliability insurance contracts with consumers, and the execution of availability contracts with generators, subject to prudential requirements to maintain solvency. This takes the form of an optimization problem ($INS$) outlined in equations (\ref{obj})-(\ref{all_o}), and described in detail thereafter. 
\begin{equation}\label{obj}
	\max_{V} (1-\beta_i)\sum_{\omega \in \Omega} \pi_{\omega} \Psi^i_{\omega} + \beta_i \tilde{\mathit{c}}^i
\end{equation}
where $\mathit{V} = \{ \overline{P_g^{N}}, \varrho_{\omega}, \gamma_{d,t,\omega},z, Q^D_d \}$, and subject to:
\begin{multline}\label{profit}
	\Psi^i_{\omega} = \sum_{d \in \mathcal{D}} C^P_d {\widehat{{P}_{d}^{D}}} Q^D_d - \sum_{t\in\mathcal{T}} \sum_{d \in \mathcal{D}} C^{sh}_d \gamma_{g,t,\omega} Q^D_d \\ - \sum_{g\in\mathcal{N}}  \overline{P^{N}_{g}} (C^{F}_g + \zeta_g C^{I}_g ), \ \forall \omega \in \Omega
\end{multline}
\begin{align}
	\sum_{\omega \in \Omega} \pi_{\omega} \Psi^i_{\omega} \geq 0 \label{aveo}  \\	
	0 \leq \gamma_{d,t,\omega} \geq \hat{p}^{sh}_{d,t,\omega} - \sum_{g\in\mathcal{N}} \overline{P^{N}_{g}}, \ \forall  d \in \mathcal{D}, t\in\mathcal{T}, \omega \in \Omega \label{gamma}\\
	\overline{P_g^{N}} \geq 0, Q^D_d \in \{0,1\} \label{triv}\\
	(\ref{cvarcon})-(\ref{rm}) \label{all_o}
\end{align}

The insurer's objective in equation (\ref{obj}) is to maximise utility, defined as a mean-CVAR measure \cite{Mays2019} where the parameter $\beta_i$ (ranging between 0 and 1) weights expected profits against CVAR based on the insurer's  preferences and $\Psi^i_{\omega}$ is the profit of the IOLR for scenario $\omega \in \Omega$, as defined in (\ref{profit}).  The first term represents premium income which is the product of the insurance premium $C^P_d$ charged to the strip of demand $d$ (quantified in terms of \$ per MW of peak demand), peak demand $\widehat{{P}_{d}^{D}}$ (in MW) and $Q^D_d$, a binary decision variable that reflects whether a reliability insurance contract is executed with demand $d \in \mathcal{D}$. The second term represents insurance compensation payouts as the product of the compensation value specified in the reliability insurance contract $C^{sh}_d$ (in \$ per MWh), $\gamma_{g,t,\omega}$ the ex-post demand curtailment (defined below) and $Q^D_d$. The third term represents payments relating to availability contracts executed with generators, which is the product of $\overline{P^{N}_{g}}$, the decision variable representing the generation capacity contracted with the insurer $g \in \mathcal{N}$, and an availability payment calculated as the sum of the generators fixed costs $C^{F}_g$ and discounted investment costs $\zeta_g C^{I}_g$. The discount factor $\zeta_g$ is calculated as an annuity based on an assumed asset life and the weighted average cost of capital for the generator. In order to eliminate the impact of financing arbitrage and focus upon contractual decision making, we make the simplifying assumption that the IOLR's investment income offsets debt service.  Nevertheless, if desired both can readily be incorporated into the model as they are fixed parameters based on the insurer's capital structure and investment portfolio.

Constraint (\ref{aveo}) ensures that the insurer is at least profitable in expectation across all scenarios. Constraint (\ref{gamma}) specifies how incremental capacity contracted by the IOLR can improve its financial position. The ex-ante demand curtailment $\hat{p}^{sh}_{d,t,\omega}$, for the set of all demand $d \in \mathcal{D}$ and time $t \in \mathcal{T}$, is the demand curtailment resulting from the market equilibrium (fixed parameters). The IOLR can contract with capacity $\overline{P^{N}_{g}}$ to reduce $\hat{p}^{sh}_{d,t,\omega}$. The resulting demand curtailment, termed as the ex-post demand curtailment $\gamma_{d,t,\omega}$ represents the curtailment after incorporating additional capacity contracted by the IOLR. This is specified as $\max(0,\hat{p}^{sh}_{d,t,\omega} - \sum_{g\in\mathcal{N}} \overline{P^{N}_{g}})$ to ensure the optimization does not allow negative curtailment, for which a linear reformulation is provided in equation (\ref{gamma}). The one bilinear term in the formulation $\gamma_{g,t,\omega} Q^D_d$ is a product of a binary variable and a continuous variable. As such, it can be represented equivalently by its McCormick envelope \cite{McCormick1976}, which results in a set of mixed integer constraints. The resultant optimization is a  mixed integer linear program (MILP) which can be solved to global optimality by modern off-the-shelf commercial solvers \cite{GurobiOptimization}.

\subsection{Decision-making framework for independent generators}

The market design for reliability insurance operates together with an electricity spot market. Independent generators are those that choose to build generation capacity based on spot market revenue alone (without receiving any availability payments from the IOLR). Hence this section develops the decision-making framework for an independent generator. The proposed approach captures the interaction between an independent generator and the electricity spot market.

We model the decision-making problem of an independent generator as a bi-level problem, where the generator will decide to invest in new power plants and retire existing ones while anticipating its profits from the electricity spot market. Here, the upper level seeks to maximise the utility resulting from the agent's investment decisions. These are constrained by spot market clearing outcomes across the set of scenarios $\omega \in \Omega$ modelled at the lower level (for avoidance of doubt these are the same scenarios observed by the IOLR).

Our modelling framework builds upon the approach of \cite{Aderhold2011}, which describes a bi-level model for generation capacity expansion. We make two modifications of note. First we extend the model from a single-scenario model to a multi-scenario model that incorporates generator risk-aversion. The risk measure chosen is a convex combination of the generator's expected profit and CVAR of the profit \cite{Mays2019}. Second we introduce constraints which reflect the financeability of generation given investor return and risk preferences. These modifications are made with the objective of incorporating a more realistic risk framework for market participants. As with \cite{Aderhold2011}, we do not model any bilateral contracts between generators and retailers in this study, though this is noted as a future extension.    

The scenario-based formulation for a generator's CVAR ($\tilde{\mathit{c}}^G_g$) is specified in (\ref{cvarcong})-(\ref{cvarcong3}) where $\Psi^G_{g,\omega}$ represents the independent generator's profits for the set of scenarios $\omega \in \Omega$, $z^G_g$ the value-at-risk (VAR), $\alpha^G_g$ the confidence level, $\pi_\omega$ the scenario probability, $\varrho^G_{g,\omega}$ is the positive deviation between scenario profits and the VAR. Variables are superscripted by $G$ to represent generators. 
\begin{align}
	\tilde{\mathit{c}}^G_g = z^G_g - \frac{1}{1-\alpha^G_g}\sum_{\omega\in\Omega} \pi_\omega \varrho^G_{g,\omega} \label{cvarcong} \\
	z^G_g - \varrho^G_{g,\omega} \leq \Psi^G_{g,\omega}, \ \forall \omega \in \Omega \label{cvarcong2} \\
	\varrho^G_{g,\omega} \geq 0, \ \forall \omega \in \Omega \label{cvarcong3}
\end{align}
The mathematical formulation for the optimization problem $(GMP_g)$ of an independent generator $g$ is provided below. The upper-level problem represents a maximization of the generator's risk-averse utility.
\begin{align}
	\max_{\{V_{\omega},\overline{{P}^G_g}\}} (1-\beta_g) \sum_{\omega \in \Omega} \pi_{\omega} {\Psi^G_{g,\omega}} + \beta_g \tilde{\mathit{c}}_g^G  \label{U_p1}
\end{align}
subject to:
\begin{align}
	{\Psi^G_{g,\omega}} = \sum_{t \in \mathcal{T}} ( \lambda_{t,\omega} -  C^{v}_g )p^{G}_{g,t,\omega} - C^{fc}_{g} \overline{{P}^G_{g}}, \> \forall \omega \in \Omega \label{psi1} \\
	\sum_{\omega \in \Omega} \pi_{\omega} \Psi^G_{g,\omega} \geq (r^e_g(1-\kappa_g)+\kappa_g r^d_g) C_g^I \overline{{P}^G_{g}} \> \>\> \label{x1} \\
	\tilde{\mathit{c}}_g^G \geq r^{d}_g \kappa_g \overline{{P}^G_{g}} C_g^I \label{y1} \\
	(\ref{cvarcong})-(\ref{cvarcong3}) \label{cv2}  
\end{align}

The objective function (\ref{U_p1}) maximises the risk averse utility for each generator $g \in \mathcal{G}$, defined as a mean-CVAR measure \cite{Mays2019}. The parameter $\beta_g$ (ranging between 0 and 1) weights expected returns against CVAR based on generator preferences. Equation (\ref{psi1}) represents the profit ($\Psi^G_{g,\omega}$) of the generator for scenario $\omega$. The first term represents the operating margin (the spot price $\lambda_{t,\omega}$ minus variable costs $C^{v}_g$ multiplied by generation dispatch $p^{G}_{g,t,\omega}$). The second term represents fixed costs $C^{fc}_{g}$ in \$ per MW, multiplied by generation capacity $\overline{{P}^G_{g}}$. Equation (\ref{x1}) and (\ref{y1}) are financing constraints, and are preconditions for investment in a generator \cite{Simshauser2018}. 
Equation (\ref{x1}) requires expected profits for the generator to exceed the weighted average cost of capital, where $r^e_g$ is the equity cost-of-capital, $r^d_g$ is the debt cost of capital, $\kappa_g$ is the gearing ratio (ranging between 0 and 1) and $C_g^I$ is the investment cost in \$ per MW. Equation (\ref{y1}) requires the CVAR exceed the debt service costs.  This ensures that in a downside scenario the generator is able to meet its debt obligations.
  
The lower level models represents the clearing of the electricity spot market under scenarios $\omega \in \Omega$ but must also give effect to the priority curtailment scheme, which prioritizes demand curtailment based on insurance contracts. In this paper, we assume a copper plate network such that the problem can be focused upon the impact of generation capacity on reliability.
\begin{multline}
	\lambda_{t,\omega}, p^{G}_{g,t,\omega} \in \arg \min_{V_{\omega}} \sum_{t \in \mathcal{T}} \sum_{g \in \mathcal{G}} C^{vc}_{g} p^{G}_{g,t,\omega} \\+ \sum_{t \in \mathcal{T}} \sum_{d \in \mathcal{D}} C^{sh}_d p^{sh}_{d,t,\omega}, \forall \omega \in \Omega
	\label{LL_lower} 	
\end{multline}
where ${V_{\omega}} = \{p^{G}_{g,t,\omega},p^{sh}_{d,t,\omega}\}$ and subject to:- 
\begin{align}
	\sum_{d \in \mathcal{D}} (\overline{{P}^D_{d,t,\omega}} - p^{sh}_{d,t,\omega}) = \sum_{g \in \mathcal{G}} p^{G}_{g,t,\omega}, \ \forall t \in \mathcal{T}, [\lambda_{t,\omega}] \label{eb} \\
	0 \leq p^{G}_{g,t,\omega} \leq  \overline{{P}^G_{g}} {A}^G_{g,t,\omega},\ \forall g \in \mathcal{G},t \in \mathcal{T}, [\underline{\mu^{G}_{g,t,\omega}},\overline{\mu^{G}_{g,t,\omega}}] \label{max_gen}\\
	0 \leq p^{sh}_{d,t,\omega}  \leq \overline{{P}^D_{d,t,\omega}},\ \forall d \in \mathcal{D},t \in \mathcal{T},
	[\underline{\mu^{sh}_{d,t,\omega}},\overline{\mu^{sh}_{d,t,\omega}}]  \label{short_min}
\end{align}

The objective function (\ref{LL_lower}) represents an economic merit-order dispatch that minimises system costs.  The first term minimises the total cost of generation for all generators $g \in \mathcal{G}$ as represented by the product of generation dispatch $p^{G}_{g,t,\omega}$ and generation variable cost $C^{vc}_{g}$. The second term minimises curtailment costs for all demand $d \in \mathcal{D}$ as represented by the product of demand curtailment $p^{sh}_{d,t,\omega}$ and the VOLL, as represented by the insurance compensation value $C^{sh}_d$.

The lower level constraints are typical of an economic dispatch. Equation (\ref{eb}) ensures power balance where the sum of generation and demand curtailment is equivalent to demand $\overline{{P}^D_{d,t,\omega}}$. Equation (\ref{max_gen}) ensures that generation dispatch is positive but below the maximum available generation limit based on the product of maximum capacity $\overline{{P}^G_{g}}$ and availability parameter ${A}^G_{g,t,\omega}$. The dual variables of each constraint are shown in square brackets.

As the lower level program is a linear program, the bilevel model can be recast as a single program by using the first order necessary and sufficient Karush-Kuhn-Tucker (KKT) conditions of the lower level problem \cite{Boyd2004a}. A more efficient representation can be derived by exploiting the property that the complementarity slackness conditions hold if and only if the strong duality condition holds \cite{Savelli2020}. As a consequence, the linear lower level problem can be equivalently represented within the bilevel program by using its primal constraints, dual constraints and the strong duality constraint.

The dual constraints for the lower level problems across the set $\omega \in \Omega$ are set out in equations (\ref{dual1})-(\ref{dual2}) and the strong duality constraint which equates the primal and dual objectives is set out in (\ref{str_dual}):

\begin{equation}
	C^{vc}_{g} - \lambda_{t,\omega} + \underline{\mu^{G}_{g,t,\omega}} - \overline{\mu^{G}_{g,t,\omega}} = 0,\forall g \in \mathcal{G},t \in \mathcal{T},[p^{G}_{g,t,\omega}] \label{dual1} 
\end{equation}
\begin{equation}
	C^{sh}_d - \lambda_{t,\omega} + \underline{\mu^{sh}_{d,t,\omega}}-\overline{\mu^{sh}_{d,t,\omega}} = 0,  \forall d \in \mathcal{D}, t \in \mathcal{T}, [p^{sh}_{d,t,\omega}] \label{dual2} 
\end{equation}
where $\mathit{W_{\omega}} = \{\lambda_{t,\omega}, \underline{\mu^{G}_{g,t,\omega}},\overline{\mu^{G}_{g,t,\omega}},\underline{\mu^{sh}_{d,t,\omega}},\overline{\mu^{sh}_{d,t,\omega}} \}$
\begin{multline}
		\sum_{t\in \mathcal{T}} \sum_{d \in \mathcal{D}} C^{sh}_d p^{sh}_{d,t,\omega} + \sum_{t \in \mathcal{T}} \sum_{g \in \mathcal{G}} C^{vc}_{g} p^{G}_{g,t,\omega} = \\ \sum_{t \in \mathcal{T}}  \sum_{d \in \mathcal{D}} \lambda_{t,\omega} \overline{{P}^D_{d,t,\omega}} 
		- \sum_{d \in \mathcal{D}} \overline{{P}^D_{d,t,\omega}} \overline{\mu^{sh}_{d,t,\omega}}  
		-  \sum_{g \in \mathcal{G}} \overline{{P}^G_{g}} {A}^g_{g,t,\omega} \overline{\mu^{G}_{g,t,\omega}} \label{str_dual}
\end{multline}	
Thus the bi-level problem introduced above can be recast into the following single equivalent optimization program:
\begin{equation}\label{SL}
	\text{(\ref{U_p1}): Upper level objective function}
\end{equation}
subject to:
\begin{align}
	(\text{\ref{psi1}) - (\ref{cv2}): Upper level primal constraints}  \label{UL_p_c} \\
	(\text{\ref{eb}) - (\ref{short_min}): Lower level primal constraints}  \label{LL_p_c} \\
	(\text{\ref{dual1}) - (\ref{dual2}): Lower level dual constraints}  \label{LL_d_c}  \\
	\text{(\ref{str_dual}): Strong duality constraint} \label{LL_sd} 
\end{align}

This single level problem is a non-convex quadratic program due to the presence of bilinear terms. There are two types of bilinear terms present: 
\begin{enumerate}
	\item  The product of two lower level continuous variables $\lambda_{t,\omega} p^{G}_{g,t,\omega}$ in (\ref{psi1})
	\item  The product of a lower level continuous variable and an upper level continuous variable, $\overline{{P}^G_{g}} \overline{\mu^{G}_{g,t,\omega}} $ in (\ref{str_dual}) 
\end{enumerate}

To remove the first bilinear term, we reformulate it based on the following lemma, with proof provided in Appendix A.

\begin{lemma} \textit{The following relationship holds at the optimum of the lower level problem:} 
	\begin{align}
		\lambda_{t,\omega}p^{G}_{g,t,\omega} = C^{vc}_{g}p^{G}_{g,t,\omega}   + \overline{{P}^G_{g}} {A}^G_{g,t,\omega}
		\overline{\mu^G_{g,t,\omega}} 
	\end{align}
\end{lemma}

This makes the second bilinear term the only remaining type of non-linearity in the model, which can be reformulated by applying a binary expansion to the variable $\overline{{P}^G_{g}}$ \cite{Wogrin2013}. 
\begin{equation}\label{binary}
	\overline{{P}^G_{g}} = \Delta_g \sum_{k=1}^{K} 2^k b_{k,g} 
\end{equation}
where $\Delta_g$ is the capacity step size, $K$ is the maximum number of steps, and $b_{k,g}$ are binary variables. The second bilinear term thus becomes $\overline{\mu^G_{g,t,\omega}} \Delta_g \sum_{k=1}^{K} 2^k b_{k,g}$. Given this term is a product of a binary variable and a continuous variable it can be represented equivalently by its McCormick envelope.

\subsection{Market equilibrium}
Each market participant is assumed to be a rational utility maximizing agent.  Each participant will seek to maximise its individual utility based on the decision making framework outlined above. An equilibrium is reached if no market participant can increase its profit by deviating unilaterally from the solution. In this paper we seek to find an equilibrium between independent generators and the IOLR, where neither the individual generator market participants nor the IOLR can profit from deviating from it.

We use a diagonalization approach to search for an equilibrium. Diagonalization solves each agent's individual decision-making problem while considering the decisions of other agents from the previous iteration \cite{StevenA.GabrielAntonioJ.ConejoJ.DavidFullerBenjaminF.Hobbs}. The diagonalization process terminates when the decision of each agent does not deviate from the last iteration.

The approach taken in this paper, described in Algorithm \ref{algo}, is similar in concept to \cite{Aderhold2011} but modified to take account of the decisions of the IOLR. Algorithm \ref{algo} commences with an initialisation of problems $(INS)$ and $(GMP_g)$.  The algorithm then iterates across independent generators to find an equilibrium between independent generators (lines 3-7). Each generator solves its individual decision-making problem while fixing the decisions of other generators to the values from the previous iteration. A \textit{spot market} equilibrium is reached when no independent generators seek to deviate from their decisions from the previous iteration. The lost load outcomes from the spot market equilibrium are fed as an input into the IOLR decision making problem to find the optimal execution of reliability insurance contracts and generator availability contracts that maximise IOLR utility (lines 8-9). The generation contracted by the IOLR is then added as additional generators in the spot market (bidding their variable costs) (line 10). The diagonalization process then repeats between the generators to find a new spot market equilibrium with the results fed back into the IOLR decision making problem. The process continues until the \textit{spot market equilibrium} outcomes and the IOLR decision-making outcomes do not change from the prior iteration. To model an \textit{energy only market} design a simple modification is made to the algorithm by adding the constraint $Q^D_d = 0, \forall d \in \mathcal{D}$ to ($INS$) thereby ensuring that no insurance contracts are executed. 

\begin{algorithm*}
	\SetAlgoLined
	\SetKwInOut{Input}{input}\SetKwInOut{Output}{output}
	\Input{Initial instance of problems ($GMP_g$) and ($INS$)}
	\Output{Equilibrium solution}
	initialization: set $\epsilon_i,\epsilon_{j}, \epsilon_k$, iteration counts $m,n$\;
	\While{$\max_{d\in\mathcal{D}}|Q^D_{d,(m)}-Q^D_{d,(m-1)}|>\epsilon_i \wedge \max_{g \in \mathcal{N}}|\overline{P^{N}}_{g,(m)}-\overline{P^{N}}_{g,(m-1)}|>\epsilon_j$}{
			\While{$\max_{g \in \mathcal{\mathcal{G}}}|\overline{{P}^G}_{g,(n)} -\overline{{P}^G}_{g,(n-1)}|>\epsilon_k$}{
			\For{$g \in \mathcal{G}$}{solve ($GMP_g$)}
			
		}
		$\hat{p}^{sh}_{d,t,\omega} \leftarrow p^{sh}_{d,t,\omega,(n)} \forall d \in \mathcal{D},t \in \mathcal{T}, \omega \in \Omega$\\ 
		solve ($INS$)\\
		$\mathcal{G} \leftarrow \{\mathcal{G} \cup \mathcal{N}\} $
		}
	return
	
		\caption{Modified diagonalization algorithm to find market and insurance equilibrium}
		\label{algo}
	\end{algorithm*}

Each run of the algorithm was tested against a range of starting conditions. In all cases, except for one the algorithm converged to the same equilibria under a range of different starting conditions.

\section{Case Study}

We evaluate the insurance mechanism design on a case study based on the South Australian system. The parameters are chosen to best illustrate the operation of the market design, rather than to recreate or predict market outcomes.

VRE is assumed to comprise wind with generation capacity sized as a percentage of total annual system electricity demand. Projections for wind availability and demand are drawn from \cite{AustralianEnergyMarketOperatorAEMO2020} across 20 annual scenarios with 17,520 dispatch intervals each (i.e. every half hour).  We set a Renewable Portfolio Standard (RPS) of 40\% of total annual system electricity demand, amounting to VRE capacity of 1620 MW.   

\begin{figure}[thpb]
	\centering
	\includegraphics[height=0.60\columnwidth]{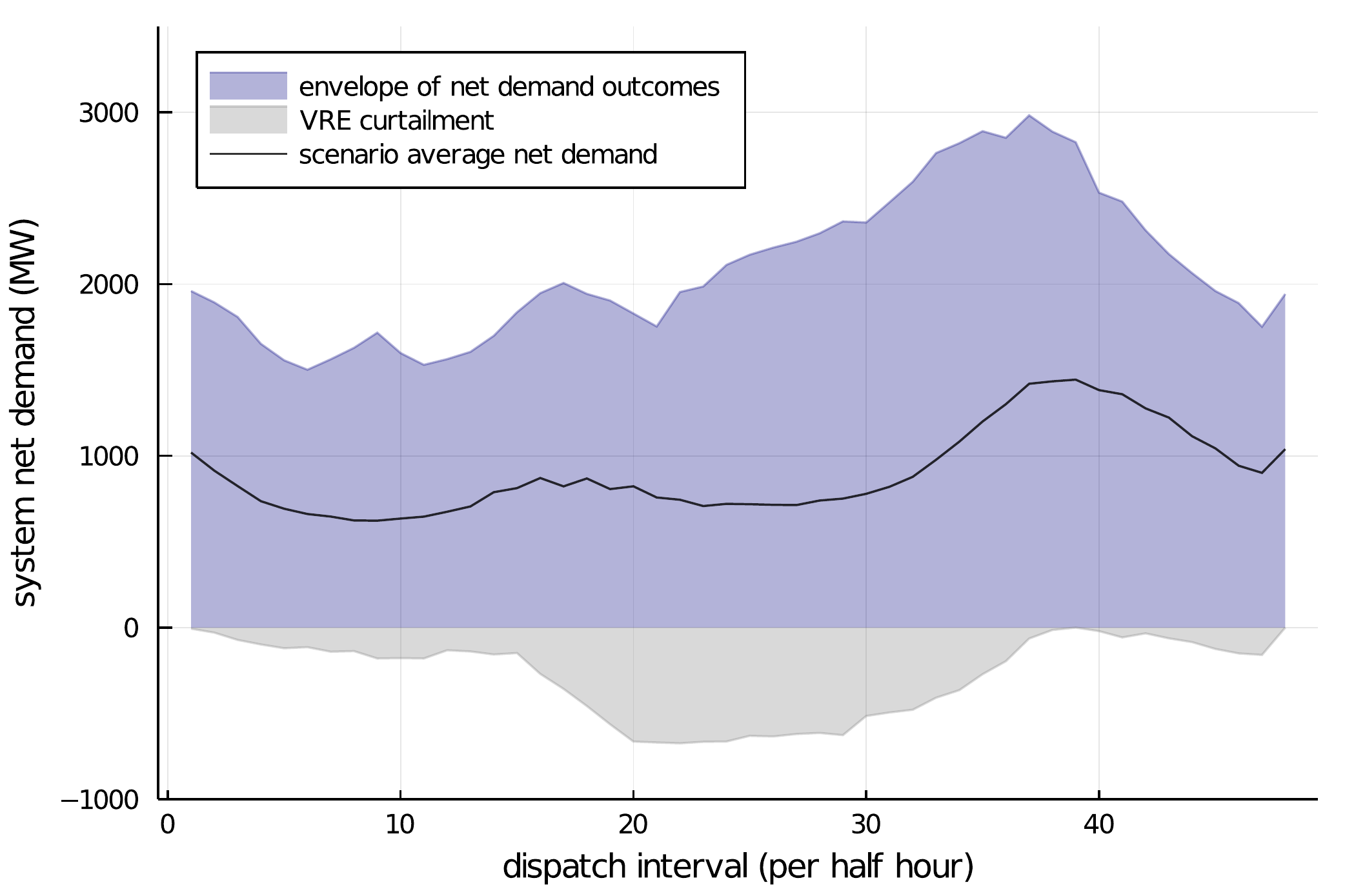}
%	\includegraphics[height=1.0\columnwidth]{Figs/Mkt_design.png}
	%	\begin{tabular}{cc}     
	%	\end{tabular}
	\caption{Net demand on an intra-day basis. Black line indicates scenario-weighted average intra-day net demand. The purple area is an envelope, for each dispatch interval, of the range of maximum and minimum net demand outcomes across all scenarios. The grey shaded region within the envelope indicates that in some scenarios VRE generation is greater than demand and needs to be curtailed.}
	\label{fig:net_demand}
\end{figure}

Twelve representative days for demand and VRE generation are selected from each of the 20 scenarios using a Ward hierarchical clustering algorithm \cite{Liu2018}.  The peak demand across all of the scenarios is 3343 MW, and the minimum demand is 551 MW. Net demand is defined as demand minus VRE generation. Fig. \ref{fig:net_demand} shows scenario-weighted average intra-day net demand and an envelope of net demand outcomes across the scenarios. The peak net demand for the system across all scenarios is 2983 MW. The grey shaded region within the envelope indicates where VRE may need to be curtailed.   
Each generator can choose to build capacity of a particular generation technology based on risk and return preferences.  Three gas-fired dispatchable generation technologies are considered. Fixed and variable costs and heat rates for the technologies are based on \cite{Administration2020}. Capital structure assumptions are based on a standard project financing with costs of debt and equity and gearing consistent with \cite{Lazard2020}. Nine independent generators are assumed to exist (three of each technology) with a generation capacity step size of 25 MW. Each participant is assumed to have an equal preference between the maximization of the scenario weighted average profits and the CVAR risk measure (i.e a $\beta_g$ of 0.5) with a confidence limit of 95\% for CVAR (i.e. $\alpha^G_g$ of 0.95). 
For the purposes of modelling a demand prioritization scheme, total system demand is segregated into two equal strips with the first strip considered as essential demand (highest priority, or last to be curtailed) and the second strip considered as non-essential demand (lowest priority, or first to be curtailed) with insurance compensation values set at \$15,000 per MWh and \$7,500 per MWh respectively. The IOLR is assumed to be capitalised with \$250 million of technical reserves with the $\alpha^i$ for the CVAR risk measure set at 0.995, consistent with international insurer solvency standards \cite{Eling2007}. The spot electricity market is cleared on the basis of optimal merit-order dispatch and settled on the marginal price with participants bidding on the basis of short-run marginal cost.  An administrative market price cap of \$2000 per MWh is imposed. For each case, we model two market designs: an energy-only market (EOM) and an energy-plus-insurance market (EIM). The code was written in Julia and solution obtained using Gurobi 9.1 on an Intel Core i7 (9th-Gen) 2.60 GHz CPU 16GB RAM. We set an optimality gap of 0.1\% for solving each optimization.

\begin{figure}[thpb]
	\centering
	\begin{tabular}{cc}     
		\includegraphics[height=0.65\columnwidth]{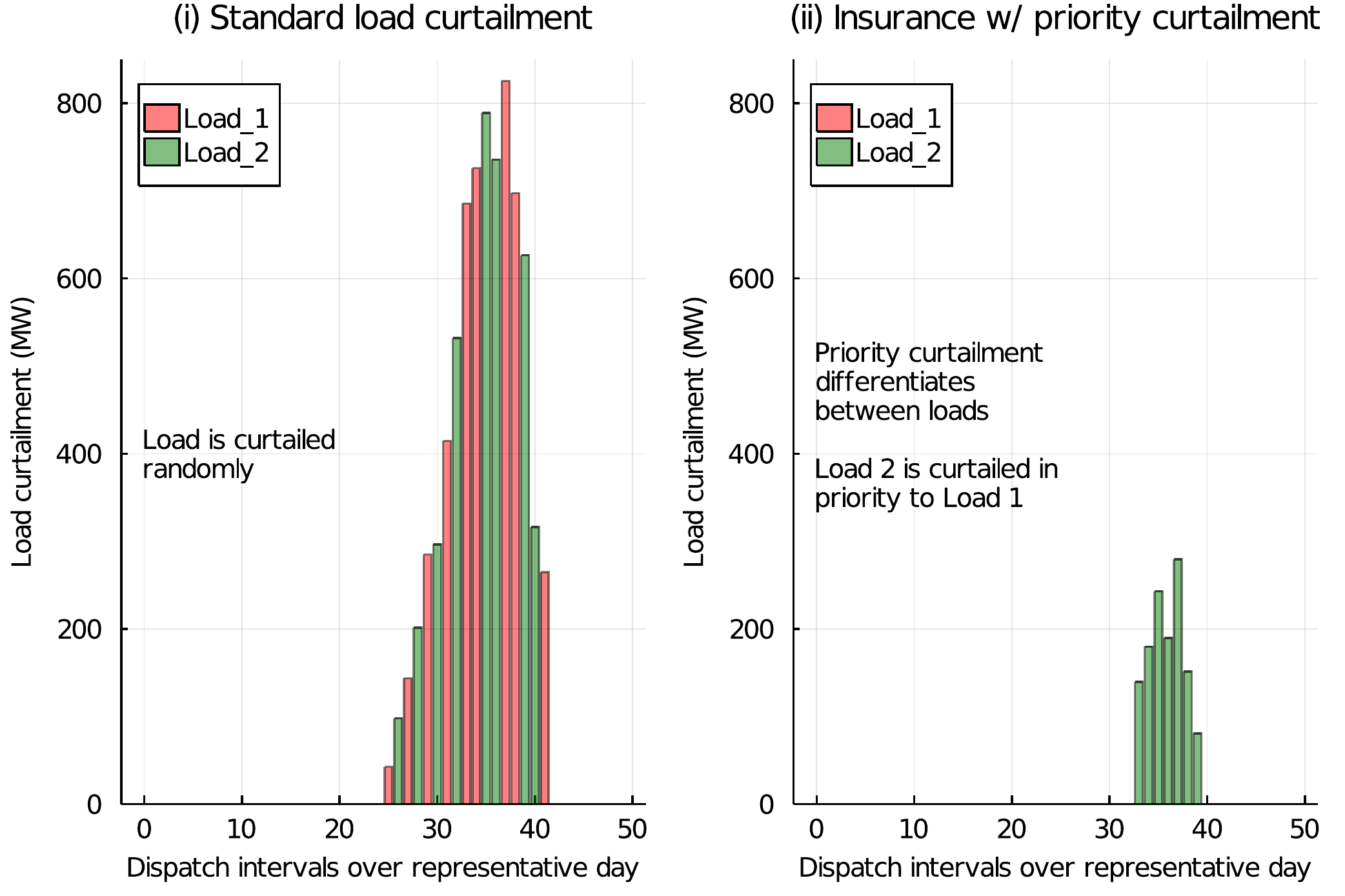} 
	\end{tabular}
	\caption{Lost load outcomes under (i) EOM with market price cap (where load is curtailed randomly) and (ii) an EIM with a priority load curtailment scheme (where load is curtailed in order of lowest insurance compensation value).  The example provided is of lost load outcomes for scenario 20, representative day 1.  Load curtailment is reduced in quantum and duration, due to incremental IOLR-supported capacity, and non-essential load is curtailed first due to the priority curtailment scheme}
	\label{fig:prior_sch}
\end{figure}

Figure \ref{fig:prior_sch} provides an example of the priority curtailment scheme in operation for a representative day. Under an EOM (the first panel), load is curtailed randomly.  This is reflective of rolling blackouts typically imposed by the system operator during extreme scarcity.  Under an EIM (second panel) with an operational priority curtailment scheme, demand is curtailed in order of priority based on the insurance compensation value specified in insurance contracts. It is observed that the prioritization of demand means that only non-essential demand is required to be curtailed during the representative day shown, allowing for uninterrupted essential demand.

\begin{table}[h]
	\caption{EOM and EIM equilibrium outcomes under 40\% RPS}
			\renewcommand\arraystretch{1.3}
	\begin{center}
		\begin{tabular}{ccc}
			\hline
			Market design & EOM & EIM\\
			\hline
			\textbf{Total capacity} (MW) & \textbf{2100} & \textbf{2571} \\
			\textit{Independent generation} & 2100 & 0 \\
			\textit{IOLR contracted generation} & 0 & 2571 \\
			\hline
			Annual load lost (\%) mean & & \\
			\textit{Demand strip 1} & 0.112 & 0.000\\
			\textit{Demand strip 2} & 0.117 & 0.005\\
			Annual load lost (\%) P95 & & \\
			\textit{Demand strip 1} & 0.450 & 0.00\\
			\textit{Demand strip 2} & 0.679 & 0.049\\			
			Value of annual load lost (\$m) mean & 161.4 & 2.2\\
			Value of annual load lost (\$m) P95 & 765.9 & 23.1\\
			\hline
			IOLR profit (\$m) mean & - & 14.7\\
			IOLR profit (\$m) CVAR & - & -250.0\\
			\hline
			\textbf{Consumer costs (\$m)} & \textbf{913.6} & \textbf{787.7} \\
			\textit{Energy} & 913.6 & 662.3\\
			\textit{Insurance} & 0 & 125.4\\
			\hline \hline
		\end{tabular}
	\end{center}
	\label{tab:res_table}
\end{table}

\begin{figure}[thpb]
	\centering
	\includegraphics[height=0.65\columnwidth]{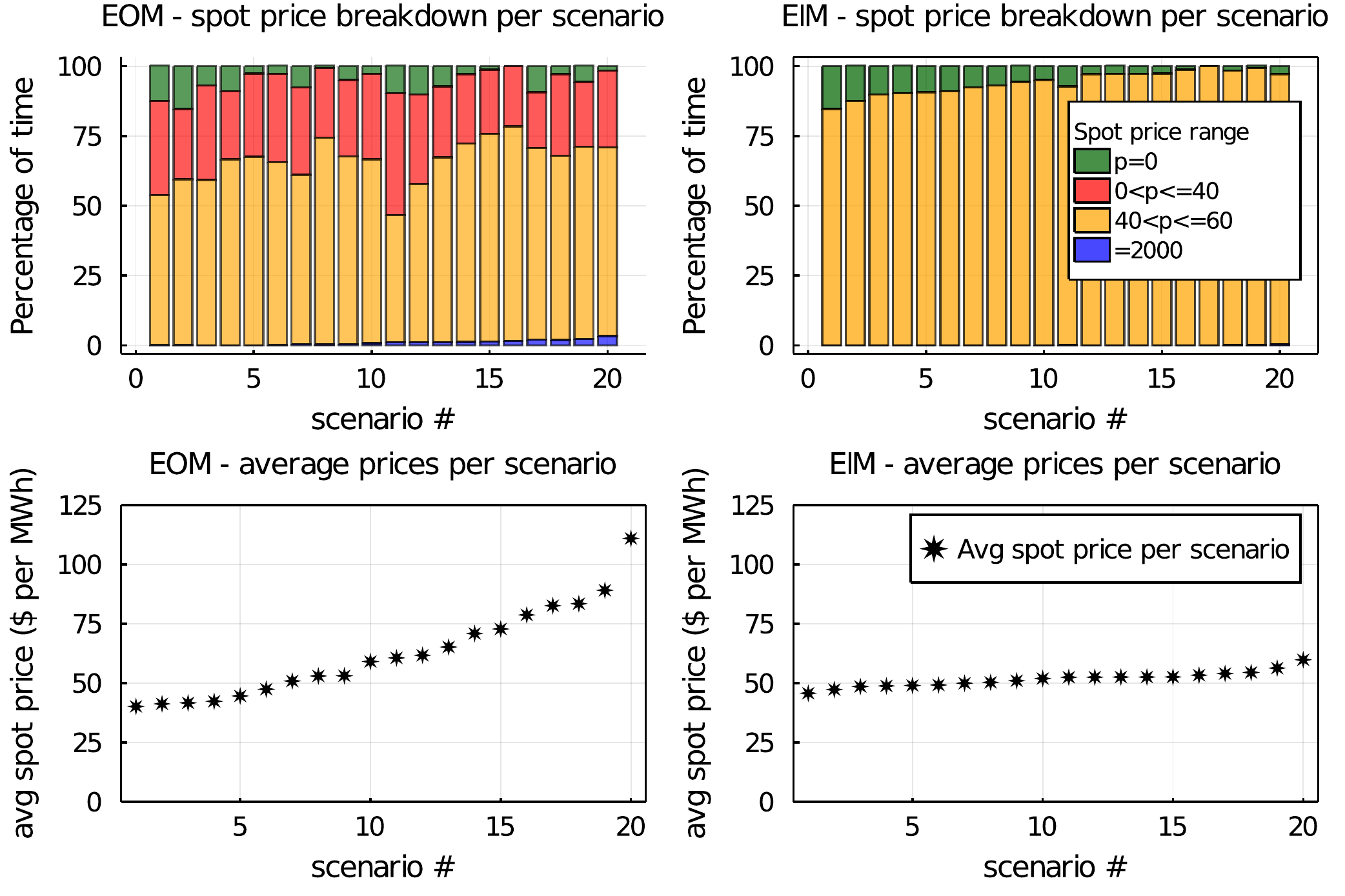} 
	\caption{Spot price outcomes under EOM and EIM designs. The top row presents a breakdown of spot prices by duration under each scenario, where the x-axis represents scenarios ($\omega \in \Omega$) and the y-axis represents the duration for which the spot price settles at various levels. The bottom row represents average spot prices for each scenario $\omega \in \Omega$.}
	\label{fig:spot_price_setters}
\end{figure}

In Table \ref{tab:res_table} we present a comparison of outcomes from an equilibrium solution of the EOM and an equilibrium solution of the EIM.  In each case, the same equilibrium was found for the case under consideration when tested against a range of starting conditions. In the EOM case, independent generation capacity built by participants is equivalent to 2025 MW. This results in annual average load curtailment of 0.112\% and 0.117\% on a scenario-weighted basis for both essential and non-essential demand respectively. For a downside case based on a 95\% scenario probability of exceedance (P95), lost load amounts to 0.45\% of essential demand and 0.679\% of non-essential demand. The cost of lost load equates to \$161 million over the year on average, and \$766 million in the P95 case. On the imposition of an insurance overlay in the EIM design, it is observed that the IOLR is incentivised to sign contracts with generation capacity of 2571 MW, exceeding the total capacity of the EOM case. This is sufficient to eliminate curtailment for essential demand and reduce curtailment for non-essential demand to 0.049\% .  The value of load lost across the system is reduced to \$2.2 million on expected value basis and \$23.1 million in the P95 case. An interesting outcome of the imposition of the insurance mechanism is that all dispatchable generation capacity that exists in the market is supported by IOLR contracts. As new supply (supported by availability contracts) enters the market, it reduces the spot price and hence revenue opportunity for remaining independent generation. This incentivises the remaining independent generators to strategically reduce capacity to maximise utility, to the detriment of system reliability. The IOLR counters this effect by supporting additional generation and this continues iteratively. An interpretation of this outcome is that it mimics the process of the commercial generation expansion, where an IOLR is incentivised to continue to monitor and make contracting decisions based on market dynamics. A related outcome is that as generation capacity is higher under an EIM design, there a reduced duration of the spot price being set at the market price cap.  This results in lower average prices (Fig. \ref{fig:spot_price_setters}) and lower total consumer costs (Table \ref{tab:res_table}).

\begin{figure}[thpb]
	\centering
	\includegraphics[height=0.67\columnwidth]{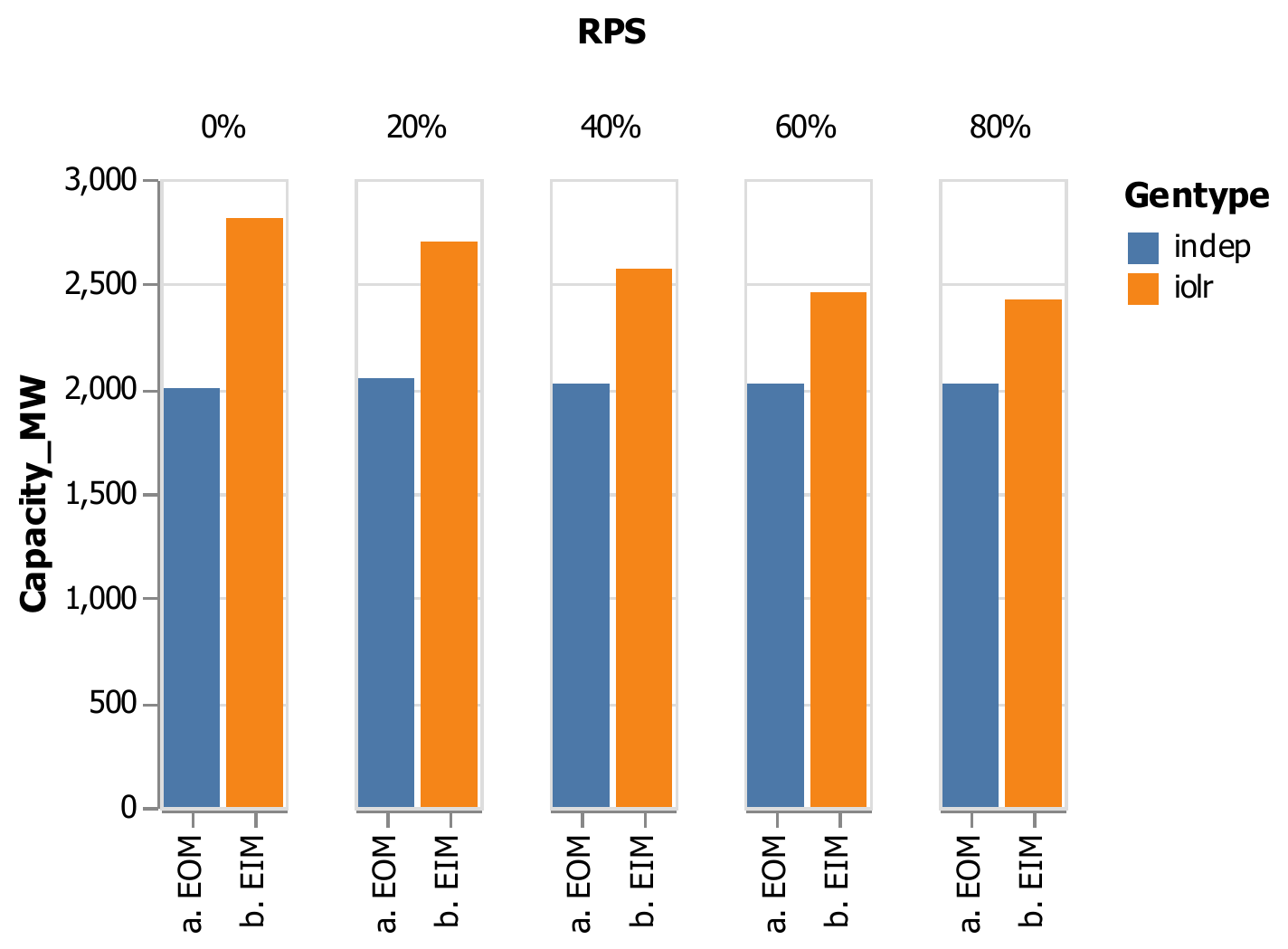} 
	\caption{Sensitivity of generation capacity for EOM and EIM market designs to different renewable portfolio standards (RPS), where VRE is sized to meet a certain percentage of annual demand. Generation capacity in an EOM is relatively insensitive to RPS levels, while generation capacity under an EIM reduces with a higher RPS.}
	\label{fig:RPS_capacity}
\end{figure}

\begin{figure}[thpb]
	\centering
	\includegraphics[height=0.7\columnwidth]{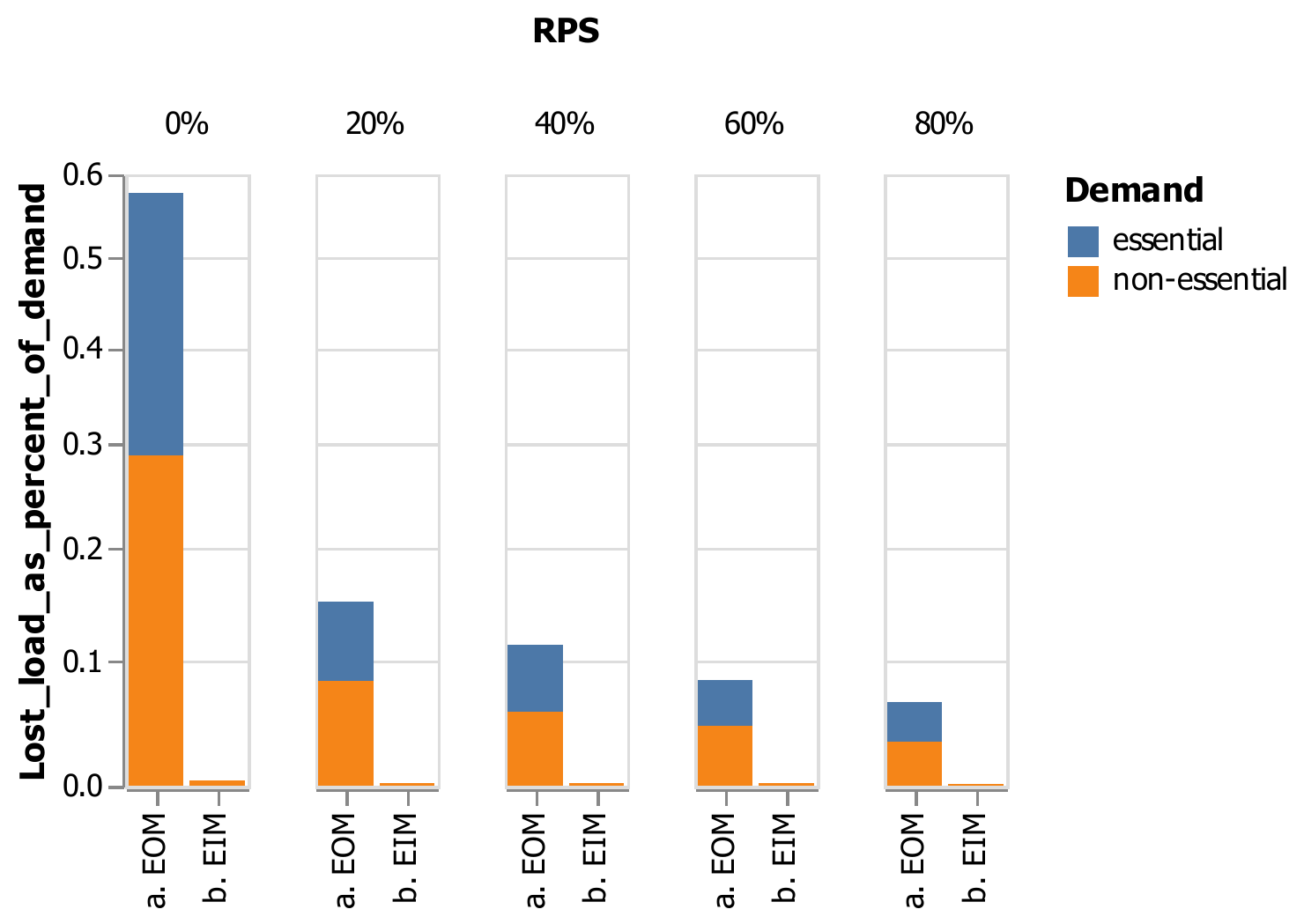} 
	\caption{Sensitivity of average lost load (as a proportion of annual demand) for EOM and EIM market designs to different renewable portfolio standards (RPS), where VRE is sized to meet a certain percentage of annual demand. Under an EOM and EIM, lost load declines with a higher RPS}
	\label{fig:RPS_ll}
\end{figure}

\begin{figure}[thpb]
	\centering
	\includegraphics[height=0.7\columnwidth]{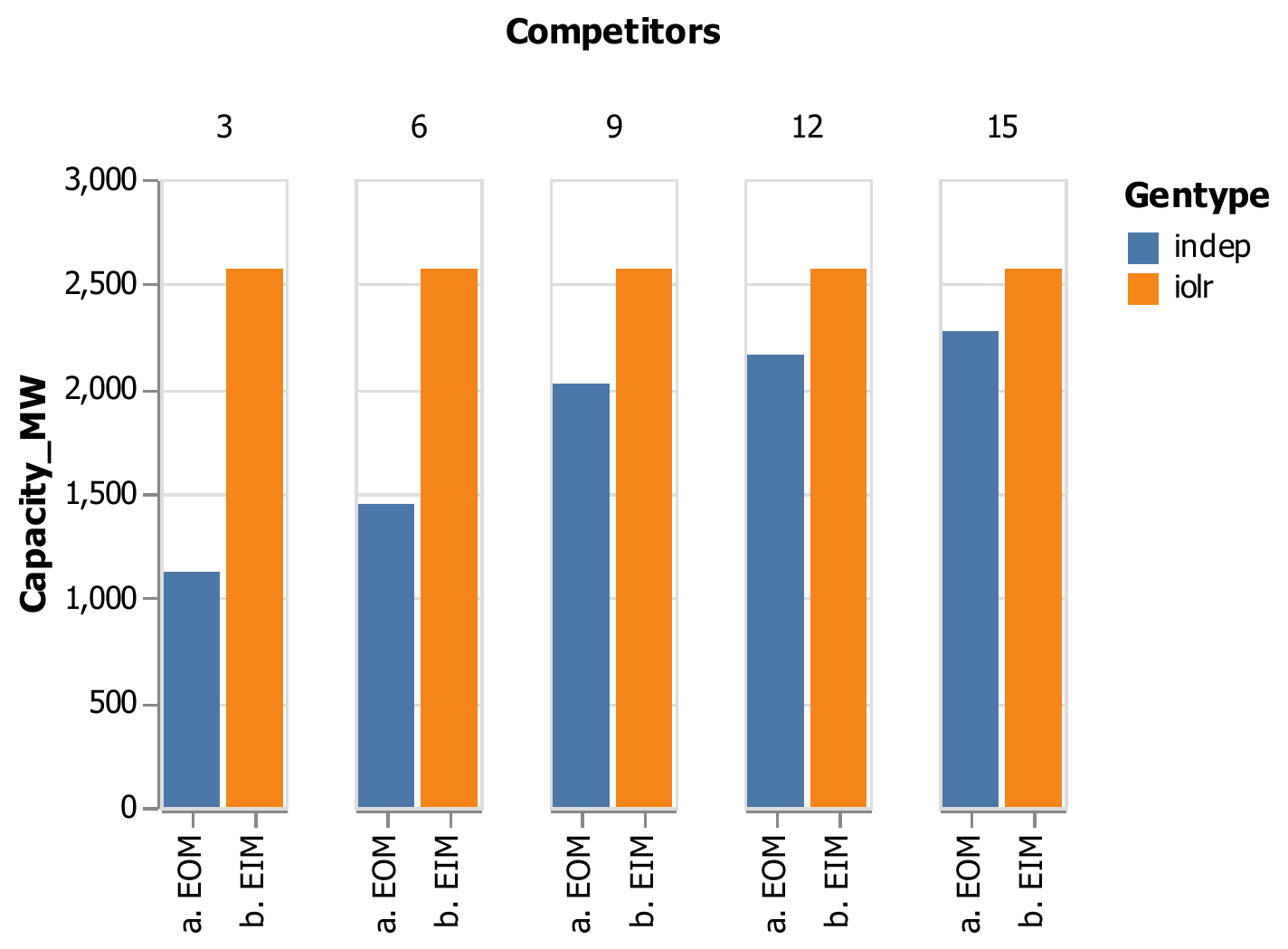} 
	\caption{Sensitivity of generation capacity for EOM and EIM market designs to the numbers of independent generators. For the EOM only independent generation is built, and for the EIM only IOLR contracted generation is built.}
	\label{fig:comp_capacity}
\end{figure}

\begin{figure}[thpb]
	\centering
	\includegraphics[height=0.7\columnwidth]{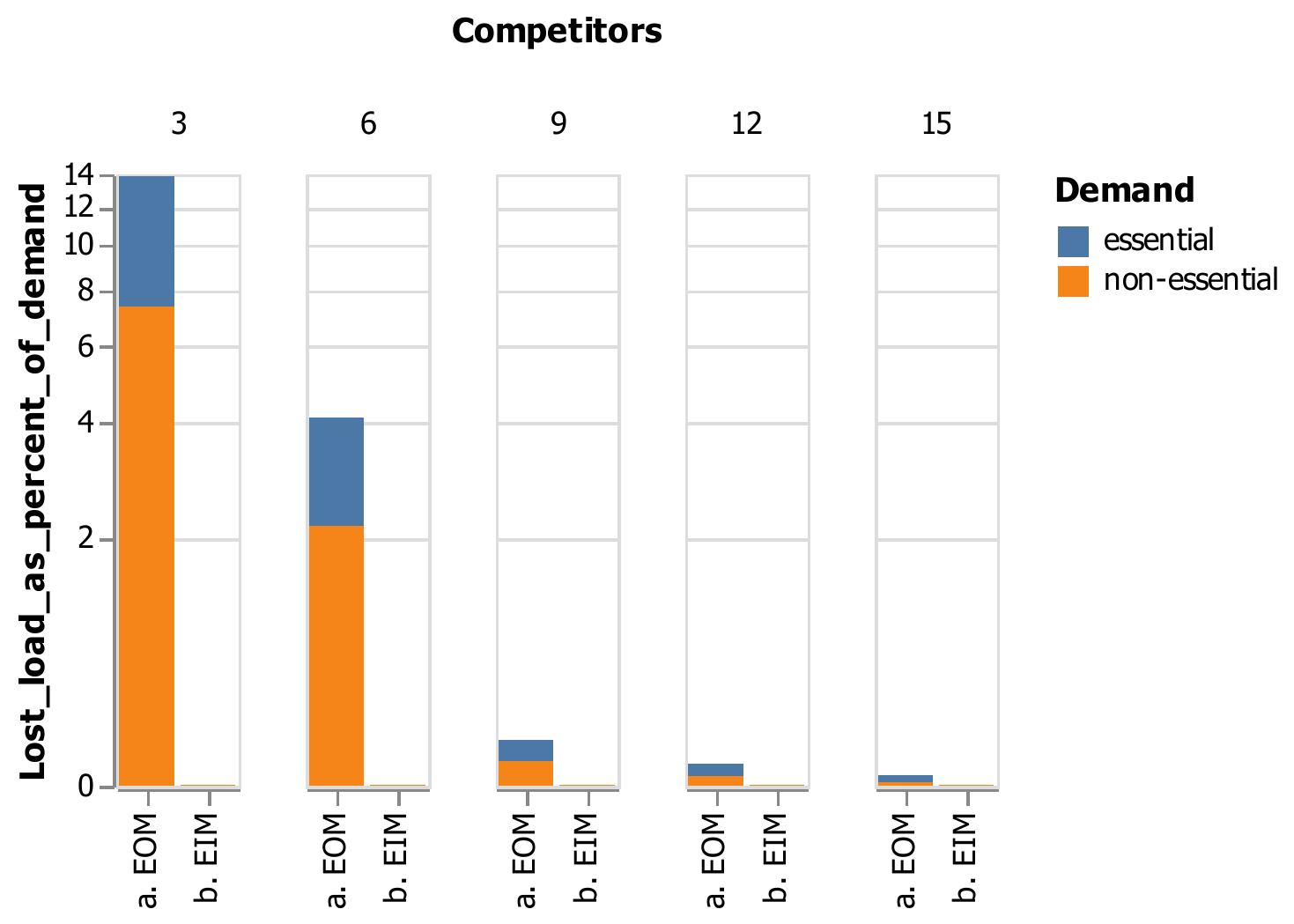} 
	\caption{Sensitivity of average lost load as a proportion of annual demand (log scale), for EOM and EIM market designs to the numbers of independent generators. For the EOM, lost load declines as the number of competitors increase, while for an EIM, lost load is insensitive to the number of competitors.}
	\label{fig:comp_ll}
\end{figure}

The sensitivity of generation capacity and reliability outcomes to different RPS levels, where VRE sized to meet 0\% to 80\% of annual demand, are shown in  Fig. \ref{fig:RPS_capacity} and Fig. \ref{fig:RPS_ll}, with the former setting out generation capacity for equilibria reached under EOM and EIM designs, and the latter setting out annualised mean lost load as a percentage of annual demand (averaged across scenarios).
Generation capacity under an EOM is relatively insensitive to RPS levels, though load outcomes are worse with a lower RPS (given reduced VRE generation in the supply mix). By contrast, the IOLR supports a higher levels of generation capacity for lower RPS.
Fig. \ref{fig:comp_capacity} and Fig. \ref{fig:comp_ll} records the sensitivity of generation capacity and reliability outcomes to the number of independent generation competitors (ranging from 3 to 15) under EOM and EIM designs. It is observed that independent generation capacity under an EOM in general increases with the number of competitors.  This reflects the importance of a competitive market dynamic for generation supply under an EOM. Under the EIM, generation capacity remains insensitive to the number of competitors in the energy market. We note that the diagonalization algorithm does not find an equilibrium for the EOM 12 generator case. However it does settle down to a reasonable range between 1975 MW and 2225 MW after 9 iterations (with an average of 2160 MW), which is the result shown in the 12 generator case for Fig. \ref{fig:comp_capacity}.
\section{Conclusions and Future Work}

In this paper, we have proposed a new reliability insurance overlay on existing energy-only markets that enables efficient generation expansion and reliability differentiation between different types of demand. Relative to an energy-only market design, the energy plus insurance design incentivises additional generation capacity that reduces the quantity of lost load at a lower overall cost for consumers. In addition the incorporation of a priority curtailment scheme also allows non-essential demand to be curtailed prior to essential demand, enabling critical services to continue during extreme scarcity. Together this suggests that an insurance-based framework can promote resource adequacy while preserving the heterogeneous value of electricity demand during scarcity. Future areas of research include investigating how the proposed design can be extended to consider (i) additional resources such as storage, (ii) regional network impacts, and (iii) a range of bilateral financial contracts between market participants.

%%%%%%%%%%%%%%%%%%%%%%%%%%%%%%%%%%%%%%%%%%%%%%%%%%%%%%%%%%%%%%%%%%%%%%%%%%%%%%%%

%%%%%%%%%%%%%%%%%%%%%%%%%%%%%%%%%%%%%%%%%%%%%%%%%%%%%%%%%%%%%%%%%%%%%%%%%%%%%%%%

%%%%%%%%%%%%%%%%%%%%%%%%%%%%%%%%%%%%%%%%%%%%%%%%%%%%%%%%%%%%%%%%%%%%%%%%%%%%%%%%
\section*{APPENDIX A: Proof of Lemma 1} \label{sec:AppA}

The non-linearity $\lambda_{t,\omega} p^g_{g,t,\omega}$ can be reformulated as follows. The dual constraint is restated in (\ref{abc}) and multiplied by $p^{g}_{g,t,\omega}$. 
\begin{equation}
	\begin{aligned}
		C^{vc}_{g} - \lambda_{t,\omega} &+ \overline{\mu_{g,t,\omega}} -\underline{\mu_{g,t,\omega}} = 0 \\
		\lambda_{t,\omega}p^{g}_{g,t,\omega} = C^{vc}_{g}p^{g}_{g,t,\omega}   &+ \overline{\mu_{g,t,\omega}}p^{g}_{g,t,\omega} -\underline{\mu_{g,t,\omega}}p^{g}_{g,t,\omega} \label{abc}
	\end{aligned}
\end{equation}

The strong duality condition (\ref{str_dual}) ensures that the complementary slackness conditions hold \cite{Savelli2020}.  Therefore using the complementary slackness conditions for (\ref{max_gen}) we obtain:
\begin{align}
	 (p^{g}_{g,t,\omega} - \overline{{P}^g}_{g} {A}^g_{g,t,\omega})  \overline{\mu^g_{g,t,\omega}} = 0 \\
	p^{g}_{g,t,\omega} \overline{\mu^g_{g,t,\omega}} = \overline{{P}^g}_{g} {A}^g_{g,t,\omega}
	\overline{\mu^g_{g,t,\omega}} \label{aaa}
\end{align}	

Using similar logic we obtain for the minimum generation condition:
\begin{align}
	\underline{\mu_{g,t,\omega}}p^{g}_{g,t,\omega} = 0 \label{abb} 
\end{align}	
By substituting (\ref{aaa}) and (\ref{abb}) into (\ref{abc}) the relation for Lemma 1 is obtained.
%%%%%%%%%%%%%%%%%%%%%%%%%%%%%%%%%%%%%%%%%%%%%%%%%%%%%%%%%%%%%%%%%%%%%%%%%%%%%%%%

\bibliography{library}{}
\bibliographystyle{ieeetr}

\end{document}